\documentclass[aps,twocolumn, floatfix, prx,longbibliography]{revtex4-1}
\usepackage{graphicx}
\DeclareGraphicsExtensions{.pdf}
\usepackage{amsmath,amssymb,bbold,bm,color}
\usepackage{float,hyperref}
\usepackage{epstopdf}
\usepackage{titlesec}
\hypersetup{colorlinks=true,urlcolor=blue}

\newcommand{\bk}{{\bm k}}

\newcommand{\br}{{\bm r}}

\newcommand{\bd}{{\bm d}}

\newcommand{\bx}{{\bm x}}

\newcommand{\by}{{\bm y}}
\newcommand{\bz}{{\bm z}}

\newcommand{\bsig}{{\bm \sigma}}

\newcommand{\bdel}{{\bm \delta}}

\newcommand{\cT}{{\cal T}}

\newcommand{\cP}{{\cal P}}
\newcommand{\cH}{{\cal H}}

\newcommand{\bee}{\begin{equation}}
  
\newcommand{\ee}{\end{equation}}

\begin{document}

\title{Unconventional superconductivity in $A$V$_2$$X_2$O family of surface altermagnets}

\author{Marcel Franz}
\affiliation{Department of Physics and Astronomy, and Quantum Matter
  Institute, University of British Columbia, Vancouver, BC, Canada V6T 1Z1}

\begin{abstract} 
Motivated by the recent discovery of superconductivity at 16.3 K in
layered oxychalcogenide Na$_{2-x}$V$_2$Se$_2$O we investigate pairing
instabilities in the broader family of layered materials composed of
V$_2$O planes, believed to exhibit
altermagnetic  order in their monolayer form. Even though the bulk
family members KV$_2$Se$_2$O and Rb$_{1-\delta}$V$_{2}$Te$_{2}$O
are likely conventional antiferromagnets that show only surface altermagnetism, our
analysis predicts exotic equal-spin triplet superconductivity as the
dominant pairing instability in these materials. This is a consequence
of their unique magnetic and sublattice structure that renders
electron bands incompatible with conventional spin-singlet
pairing. The predicted triplet superconducting phases are
topologically non-trivial and capable of supporting spin-polarized
persistent currents, properties potentially useful in technological applications.   

\end{abstract}

\date{\today}
\maketitle

{\em Introduction --} Vanadium-based oxychalcogenides
KV$_2$Se$_2$O and Rb$_{1-\delta}$V$_{2}$Te$_{2}$O were reported as
candidates for altermagnetic metals \cite{Smejkal2022a,Smejkal2022b,Mazin2022,Jungwirth2026} based on angle-resolved photoemission
(ARPES) experiments  \cite{Jiang2024, Zhang2024} that showed clear
evidence of significant spin splitting in 
their single-electron bands. Altermagnetism is thought to originate from their V$_2$O
planes arranged in the so-called ``anti-CuO$_2$'' structure with
oxygens occupying the corner sites and vanadium the bond sites of the
square lattice. When the vanadium magnetic moments order
antiferromagnetically the resulting monolayer structure lacks the combined
inversion ($\cP$) and time-reversal ($\cT$) symmetry which qualifies
it as a 2D altermagnet \cite{Sudbo2023,Fernandes2025,Kaushal2025}. 

\begin{figure}[t]
  \includegraphics[width=8.5cm]{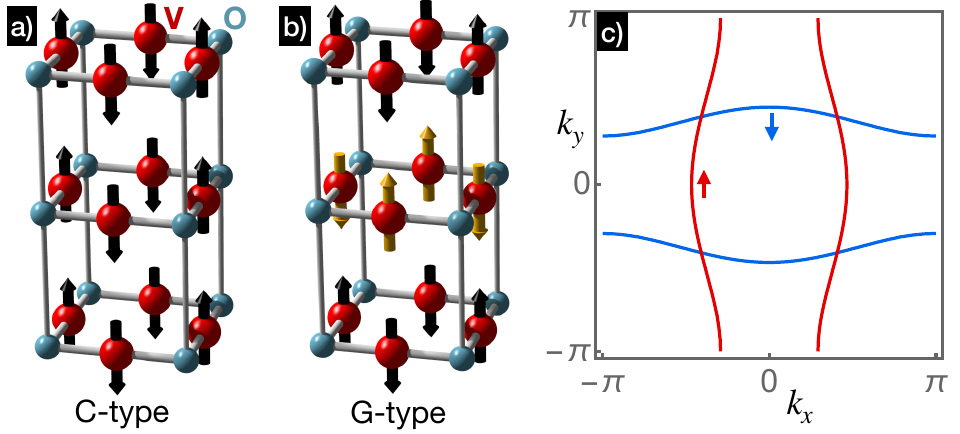}
  \caption{(a,b) Magnetic and crystal structures of vanadium-based
    oxychalcogenides. For the sake of clarity only the V$_2$O planes
    are shown. (c)  Spin split Fermi surfaces of a $d$-wave
    altermagnetic metal modeling a single V$_2$O  monolayer
    described by Hamiltonian Eq.\ \eqref{h1} for altermagnetic band spitting parameter    $\eta=0.6$ and $\mu=-1.14$. Red and blue color distinguishes up and down spin projections.}
  \label{fig1}
\end{figure}

A 3D material formed by stacking such V$_2$O layers will behave as a bulk
altermagnet provided that vanadium moments
order {\em  ferromagnetically} between the layers in a C-type magnetic structure
illustrated in Fig.\ \ref{fig1}(a). Recent neutron diffraction
experiments \cite{Sun2025,Yang2026,Xie2026} as well as first principles density functional
studies \cite{Thapa2026} however reported G-type bulk magnetic structure  Fig.\ \ref{fig1}(b)
with {\em antiferromagnetic} interplane order.
In this case the 3D
crystal acquires $\cP\cT$ symmetry with a center of inversion midway
between the planes. As a result  all bands become doubly spin-degenerate
and the system behaves as a conventional AF metal
suggesting that KV$_2$Se$_2$O
and Rb$_{1-\delta}$V$_{2}$Te$_{2}$O are not bulk altermagnets after all.

The ARPES data \cite{Jiang2024, Zhang2024}  can be reconciled with the G-type bulk magnetic
structure by invoking the concept of {\em surface altermagnetism}
\cite{Lange2026,Leeb2026}. This refers to a situation where a system
that is a bulk antiferromagnet (AF)
exhibits spin-split bands near its surfaces due to the locally broken
inversion symmetry. We recall that ARPES is a surface-sensitive technique that probes
primarily the topmost atomic layer of the crystal. An isolated V$_2$O monolayer
is a bona-fide 2D altermagnet
\cite{Sudbo2023,Fernandes2025,Kaushal2025}  and  it is
intuitively plausible that it would show 
 characteristic spin splitting even when weakly coupled to the rest
of the crystal. From the symmetry perspective, the presence of a
surface breaks the inversion symmetry, $\cP\cT$ is locally
broken, and spin-split bands become symmetry-allowed in surface-sensitive
probes.

In this Letter, we construct a minimal model of a surface
altermagnet with AF bulk relevant to the $A$V$_2$$X_2$O ($A$ = Na, K, Rb,
Cs; $X$ = Se, Te)  family of materials with G-type magnetic
structure. We show by an explicit calculation that although bulk bands 
are all doubly spin-degenerate, spin-resolved spectral functions of the
surface layers can show pronounced altermagnetic splitting, in accord with
the ARPES data 
\cite{Jiang2024, Zhang2024} and earlier theoretical work
\cite{Lange2026,Leeb2026}. Then, motivated by the 16.3 K 
superconductivity  reported in Na$_{2-x}$V$_2$Se$_2$O
\cite{Sun2026SC}, we study the superconducting (SC) 
instabilities of our model in the presence of weak attractive
interactions between electrons.

It has been
argued previously that due to their spin-split Fermi surfaces (FS) altermagnetic
metals are fundamentally  incapable of forming conventional spin-singlet
superconducting phases -- instead, the dominant instability at weak
coupling is towards the exotic equal-spin triplet state
\cite{Zhu2023,Heung2024,Leraand2025,Monkman2026,Fradkin2026}. Here we
demonstrate that even though the AF metal representing
bulk oxychalcogenides does not have spin-split FS, its
leading SC instability, remarkably, is also in the equal-spin triplet channel. As
we explain, this occurs because of the combined magnetic and sublattice
structure of  the degenerate metal with G-type magnetic ordering which
prevents formation of spin-singlet Cooper pairs. This suppression of singlet
pairing makes the otherwise normally weaker triplet instability
dominant in these materials. Hence, we predict that
Na$_{2-x}$V$_2$Se$_2$O is an exotic spin-triplet superconductor.   

{\em Model --} We describe electrons in the V$_2$O monolayer using an effective
single-band Hamiltonian for a $d$-wave altermagnet
\cite{Smejkal2022a,Smejkal2022b}
\begin{equation}\label{h1}
 h_0(\bk)=t_\bk+\eta_\bk\sigma_z,
\end{equation}
with  $ t_\bk=-2t(\cos{k_x}+\cos{k_y})$ and $\eta_\bk=-2\eta(\cos{k_x}-\cos{k_y})$.
Here, $\sigma_{a=x,y,z}$  are Pauli matrices in spin space, $t$ denotes the nearest
neighbor hopping amplitude on the square lattice. The term $\eta$ encodes the
splitting of the altermagnetic bands and breaks the time-reversal
symmetry $\cT$ represented by $i\sigma_yK$, where $K$ denotes complex
conjugation. The model remains invariant under combined $C_4$
rotation and $\cT$. For suitably chosen parameters, it produces FS
similar to those found in oxychalcogenides Fig.\ \ref{fig1}(c). 

\begin{figure}[t]
  \includegraphics[width=8.5cm]{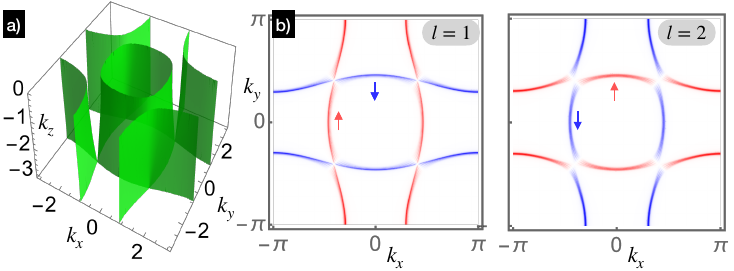}
  \caption{a) Bulk FS of the model Hamiltonian Eq.\
    \eqref{h2}. b) Layer-resolved spectral functions Eq.\
    \eqref{h3} for two topmost layers in a stack of $M_z=12$ V$_2$O
    layers with G-type magnetic order described by the same Hamiltonian.
We use $\eta=0.6$, $\mu=-1.14$, $g=0.4$ and $\delta=0.03$ but the
results look similar for a broad range of parameters and choices of $M_z$.}
  \label{fig2}
\end{figure}

We next build the 3D crystal by stacking individual monolayers. For a
G-type magnetic structure  there will be two monolayers with opposite
value of $\eta$ in a unit cell. The bulk Hamiltonian can then be
written as 
\begin{equation}\label{h2}
h(\bk)=t_\bk+\lambda_z\sigma_z\eta_\bk+\lambda_xg_\bk,
\end{equation}
where  $\lambda_a$ are Pauli matrices in the sublattice space and
$g_\bk=-2g\cos{(k_z/2)}$ describes spin-independent  interplane
electron tunneling with amplitude $g$. If we select the center of
inversion midway between two V$_2$O planes forming a unit cell then
the inversion operator $\cP$ can be 
represented by $\lambda_x$. It is then easy to see that
$h(\bk)$ respects the combined $\cP\cT$ symmetry. Because
$(\cP\cT)^2=-1$ it follows that all bands are doubly degenerate. This
can also be seen directly by diagonalizing the $4\times 4$ matrix
Hamiltonian \eqref{h2} which yields only two distinct energy eigenvalues
\begin{equation}\label{h2a}
\epsilon_{\bk\alpha}=t_\bk+\alpha\sqrt{\eta_\bk^2+g_\bk^2}, \ \ \
\alpha=\pm 1.
\end{equation}
The corresponding spin-degenerate FS is shown in Fig.\
\ref{fig2}(a). We conclude that for a  G-type magnetic ordering, there
is indeed no altermagnetic splitting in the bulk system.

The combined $\cP\cT$ symmetry implies that the {\em bulk} electron
spectral functions for the two spin projections are the same,
$A_\uparrow(\bk,\omega)=A_\downarrow(\bk,\omega)$. However, as a
surface-sensitive probe, the ARPES signal is dominated by the topmost layer
of the crystal. Hence, to model the ARPES data we consider a slab
consisting of a finite number of unit cells and define a spectral
function associated with each individual V$_2$O layer,
\begin{equation}\label{h3}
A_{l,\sigma}(\bk_\perp,\omega)=-{\rm Im}[\omega+i\delta
-\hat{h}(\bk_\perp)]^{-1}{\big |}_{ll,\sigma}.
\end{equation}
Here $\bk_\perp=(k_x,k_y)$ denotes the in-plane crystal momentum and
$\hat{h}(\bk_\perp)$ is a $4M_z\times 4M_z$ matrix Hamiltonian obtained
by recasting Eq.\ \eqref{h2} in the real-space representation along
the $z$-direction for a slab with $M_z$ layers. The subscript
$(ll,\sigma)$ in Eq.\ \eqref{h3} indicates the diagonal element of the matrix
corresponding to the layer $l$ and the spin $\sigma$.

Spectral functions of the two top layers $(l=1,2)$ in a slab with
$M_z=12$ layers are displayed in Fig.\ \ref{fig2}. These results
confirm that any probe sensitive primarily to the top layer
would detect a significant spin splitting, in accord with ARPES data
on  KV$_2$Se$_2$O and Rb$_{1-\delta}$V$_{2}$Te$_{2}$O \cite{Jiang2024,
  Zhang2024}. In fact, we find that the spectral function of 
any individual V$_2$O layer shows pronounced spin splitting, even
though, as we emphasized, the bulk spectral function
$A_\sigma(\bk,\omega)$ is fully spin-degenerate.  This unique behavior
is a hallmark of the surface altermagnet and underlies the
unconventional superconductivity discussed below.

{\em Superconducting instabilities --} We are interested in the ground
state of the magnetic metal described by the Hamiltonian Eq.\ \eqref{h2} in the
presence of weak attractive interaction between electrons. We consider
both on-site and nearest-neighbor attraction described by 
\begin{equation}\label{h4}
\cH_I=-V_0\sum_\br n_{\br\uparrow}n_{\br\downarrow}
-\sum_{\br,\bdel}V_1^\bdel n_{\br}n_{\br+\bdel}, 
\end{equation}
where $n_{\br\sigma}$ is the electron number
operator on site $\br$ with spin $\sigma$,
$n_{\br}=n_{\br\uparrow}+n_{\br\downarrow}$ and $\bdel$ extends over the
nearest neighbor vectors $(\hat\bx, \hat\by,\hat\bz)$. Positive values
of $V_0$ and $V_1^\bdel$ signify attraction.  By symmetry we
have $V_1^x=V_1^y\equiv V_1$ and for weakly coupled V$_2$O layers in
oxychalcogenides we also expect the inter-plane interaction to be much
weaker than in-plane, $V_1^z\ll V_1$.

\setlength{\tabcolsep}{6pt} 
\begin{table*}[t]
\centering 
\caption{Singlet and triplet order parameters that arise from decoupling the interaction
  Hamiltonian $\cH_I$ defined in Eq.\ \eqref{h4}. Results in the last
two columns are obtained for a stack of $M_z=24$ V$_2$O
 layers with periodic boundary conditions in all directions. 
We use $\eta=0.6$, $\mu=-1.14$, $g=0.4$ and $V_0=V_z=V_1^z=3.0$ to
compute the zero-temperature maximal gap $\Delta(0)$ and the critical
temperature $T_c$ by solving the relevant gap equations through
numerical iteration.}
\begin{tabular}{c c c c c c}
\hline\hline
 ~\  Order parameter  ~\  ~\  &\  ~\ ~\ $d^0_\bk$  or $\bd_\bk$ \  ~\  ~\
  &\  ~\ ~\ BdG matrix\  ~\ ~\
  &\  ~\ ~\ coupling ~\ ~\ &\  ~\ ~\ max gap $[t]$ ~\ ~\ &
                                                            \  ~\ ~\
                                                            $T_c [t]$
                                                            ~\ ~ \\
\hline
  on-site singlet &            $\Delta_0$      &
                                                                       $\tau_x$
                                                                       
  &  $V_0$  & $0.03625$  & $0.0159$\\
  \hline
  
in-plane singlet   & $\Delta_s(\cos{k_x}+\cos{k_y})$ & $\tau_x$  & $V_1$
                                                             &
                                                               $0.00687$
  & $0.0033$\\
 interplane singlet   & $\Delta_s^z\cos{(k_z/2)}$ & $\tau_x\lambda_x$  & $V_1^z$
                                                             &
                                                               $0.02590$
  & $0.0127$ \\

  \hline
in-plane triplet &            $\Delta_p(\sin{k_x},\pm i \sin{k_y},0)$      &
                                                                       $\tau_x\sigma_x$,
                                                                       $\tau_x\sigma_y$
  &  $V_1$  & $0.54858$ &  $0.3661$ \\

interplane triplet   & $\Delta_p^z(0,0,\sin{(k_z/2)})$ & $\tau_x\lambda_x\sigma_z$  & $V_1^z$
                                                             &
                                                               $0.02772$
                                                          &  $0.0135$\\
\hline\hline
\end{tabular}
\label{table1}
\end{table*}

To understand the nature of SC instabilities we next perform the standard mean-field
decoupling of $\cH_I$ in all possible pairing channels. As outlined in the
Supplementary Material \cite{SM} this results in 5
independent order parameters comprising on-site spin singlet and
singlet/triplet on the in-plane and inter-plane n.n.\ bonds. 
The Bogoliubov-de Gennes (BdG) Hamiltonian describing the system after
the decoupling can be written as
\begin{equation}\label{h5}
  H(\bk)=
  \begin{pmatrix}
    h(\bk)-\mu & \hat\Delta_\bk \\
    \hat\Delta_\bk^\dag & -\sigma_y h(-\bk)^*\sigma_y+\mu
    \end{pmatrix},
\end{equation}
where $\hat\Delta_\bk$ denotes the pairing matrix that contains all 5
SC orders. The lower diagonal block is the $\cT$-conjugate of $h(\bk)$
and $\mu$ denotes the chemical potential. $H(\bk)$ is an $8\times 8$
matrix in the combined spin, sublattice and Nambu space in the basis
defined as $\Psi_\bk=(\psi_\bk,i\sigma_y\psi^*_{-\bk})^T$. Here
$\psi_\bk=(a_{\bk\uparrow}, a_{\bk\downarrow},  b_{\bk\uparrow},
b_{\bk\downarrow})^ T$ and $ a^\dag_{\bk\sigma}/b^\dag_{\bk\sigma}$
create electrons with spin $\sigma$ and momentum $\bk$ on the two sublattices.

The pairing matrix can be written in the standard $d$-vector notation as 
\begin{equation}\label{h8}
 \hat\Delta_\bk=
  \begin{pmatrix}
    \bd_\bk\cdot\bsig  & d^0_\bk\\
    d^0_\bk & \bd_\bk\cdot\bsig
    \end{pmatrix}.
\end{equation}
where $d^0_\bk$ and $\bd_\bk$ encode the singlet and triplet
components of the order parameter, respectively. Their structure is
summarized in Table I. Note that in terms of in-plane triplet we will
focus on the chiral $p$-wave orbital wavefunction $\sim(p_x\pm i p_y)$
as opposed to the nematic (pure $p_x$ or $p_y$). The former produces a
fully gapped excitation spectrum and is therefore expected to be lower
in energy, which we confirm by an explicit calculation. We also remark
that in the triplet case the BdG Hamiltonian 
becomes block-diagonal in spin and hence yields two independent
$p$-wave condensates for spin-up and spin-down electrons.
%

It is
instructive to write the two diagonal blocks of $H(\bk)$ explicitly as
\begin{equation}\label{h6}
  H(\bk)^{\rm
    diag}=\tau_z(t_\bk-\mu)+\lambda_z\sigma_z\eta_\bk+\tau_z\lambda_xg_\bk, 
\end{equation}
where $\tau_a$ denotes Pauli matrices in the Nambu space. The
corresponding matrix form of various order parameters is given in the
third column of Table I.  It is easy to check that both triplet order
parameters and the interplane singlet are represented by $8\times 8$
matrices that anticommute with all the terms in the normal part of
the Hamiltonian given in Eq.\ \eqref{h6}. 
 This property ensures the canonical  BCS form of the spectrum $\pm
 E_{\bk\alpha}$ with 
\begin{equation}\label{h6a}
E_{\bk\alpha}=\sqrt{(\epsilon_{\bk\alpha}-\mu)^2+|\Delta_\bk|^2},
\end{equation}
which is fully gapped whenever $|\Delta_\bk|>0$ on the FS. The gap
function is
given as $\Delta_\bk=d^0_\bk$ for a singlet and $\Delta_\bk=|\bd_\bk|$
for a triplet.

The singlet order parameters listed in the first two lines of Table I
are associated with matrix $\tau_x$ which commutes with the magnetic
$\eta_\bk$ term in the normal Hamiltonian Eq.\ \eqref{h6}. This
results in a more complicated non-BCS form of the excitation spectrum,
which we analyze in more detail in SM \cite{SM}.

{\em Quantitative analysis --}
It is straightforward to derive the BCS gap equations for various order
parameters listed in Table I. The general form is obtained by minimizing the system free energy
\cite{SM} and reads
\begin{equation}\label{h9}
    \Delta={V\over 4 N}\sum_{\bk,\alpha}{\partial E_{\bk\alpha}\over \partial\Delta}\tanh{{\beta  E_{\bk\alpha}\over 2}}.
\end{equation}
where $\beta=1/k_BT$ and $N$ is the number of unit cells in the
crystal,  $\Delta$ represents one of the order parameters
$\{\Delta_0,\Delta_s,\Delta_s^z,\Delta_p,\Delta_p^z\}$ and $V$ denotes
the corresponding coupling listed in column 4 of Table I. As an
example, the gap equation for the interplane singlet component takes
the form
\begin{equation}\label{h10}
  \Delta_s^z={V_1^z\over 4
              N}\sum_{\bk,\alpha}{\Delta_s^z\cos^2{(k_z/2)}\over
              E_{\bk\alpha}}\tanh{{\beta  E_{\bk\alpha}\over 2}},
\end{equation}
where $E_{\bk\alpha}$ is given by Eq.\ \eqref{h6a} with
$\Delta_\bk=\Delta_s^z\cos{(k_z/2)}$.

\begin{figure}[t]
  \includegraphics[width=8.5cm]{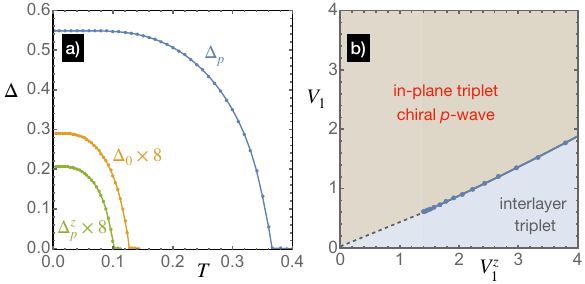}
  \caption{a) Temperature dependence of order parameters $\Delta_p$,
    $\Delta_0$ and $\Delta_p^z$
    parameters obtained by solving the relevant gap equations based on
    Eq.\ \eqref{h9}  through numerical iteration. We use the same model
    parameters as Fig.\ \ref{fig2} with $V_0=V_1=V_1^z=3.0$. Note that
    graphs for $\Delta_0$ and $\Delta_p^z$ have been scaled along both
    axes by a factor of 8 for clarity.
    b) The resulting phase diagram in $V_1^z$-$V_1$ plane obtained by
    comparing critical temperatures in the two channels. The dashed
    line is an extrapolation of numerical results: At small values of the
    interaction strength gaps become exponentially small and solving
    the gap equations becomes numerically challenging.
}
  \label{fig3}
\end{figure}

We solve the gap equations by numerical
iteration and display the typical temperature dependences of the three
largest order parameters in Fig.\ \ref{fig3}(a). The extracted values of
the maximum gap $\Delta(T=0)$ and the critical temperature $T_c$ are listed in
columns 5 and 6 of Table I. While these naturally depend on model
parameters, we find that the in-plane triplet $\Delta_p$ is
always the dominant order parameter in that it exhibits by far the
largest $T_c$ and $\Delta(0)$, provided that we assume comparable values of
all interaction parameters $V$. Mathematically, we attribute this
result to the fact that among all order parameters the chiral
$p$-wave alone is capable of producing a fully gapped state and
hence its nucleation gains the greatest amount of  condensation energy. This leads
to our main conclusion that the spin-triplet chiral $p$-wave order should
robustly prevail in these materials.

The on-site singlet $\Delta_0$ and
in-plane triplet $\Delta_p^z$ are next in line, but can be stabilized
only when the corresponding interaction parameters $V_0$ and $V_1^z$
are larger than $V_1$ by at least a factor of 2. However, we do not expect the real materials
to be in this limit. As already mentioned, we expect $V_1^z\ll V_1$ due
to the quasi-2D structure of vanadium oxychalcogenides with weakly
coupled layers. We also expect $V_0<V_1$ on account of the strong
on-site Coulomb repulsion between electrons that cannot be efficiently
screened. Hence, it is reasonable to assume that the in-plane n.n.\
interaction parameter $V_1$ is generically the largest attraction scale in the problem.   
The phase diagram in the $V_1^z$-$V_1$ plane, constructed by comparing
the relevant critical temperatures, is presented
as  Fig.\ \ref{fig3}(b). It shows that most of the phase space is
taken by the in-plane triplet $p$-wave while the interplane triplet phase occurs only 
when $V_1^z\gtrsim 2 V_1$. A similar diagram in the $V_0$-$V_1$ plane
showing competition between $\Delta_p$ and $\Delta_0$ is displayed in the SM
Fig.\ \ref{fig_s1}.

{\em Discussion --}  Theoretical analyses \cite{Lange2026,Leeb2026,Thapa2026} suggest that
vanadium-based oxychalcogenides in the $A$V$_2$$X_2$O family are
{\em surface altermagnets} characterized by spin-degenerate
electron bands in their bulk accompanied by a significant
altermagnetic spin splitting at their surfaces. This view reconciles
the ARPES data reporting strong spin splitting in surface bands
\cite{Jiang2024, Zhang2024}  with the G-type bulk
magnetic structure observed by inelastic neutron scattering \cite{Sun2025,Yang2026,Xie2026}.
Starting from a minimal model of a layered surface altermagnet we
showed that the leading superconducting instability in the presence of
weak attractive interactions in such materials is an 
exotic equal-spin triplet with chiral $p$-wave symmetry.

This result can be understood on the basis of the following
two observations. First, as
illustrated in Fig.\ \ref{fig1}(b), the $A$V$_2$$X_2$O structure can be pictured as
composed of weakly coupled V$_2$O layers. In isolation each such layer
would exhibit $d$-wave
altermagnetic order with spin-split Fermi surfaces of the type
sketched in Fig.\ \ref{fig1}(c).
Second, for weakly coupled  V$_2$O layers, we expect the dominant attractive
interaction to reside in the individual layers. Hence, the
analysis of SC instabilities must begin by considering pairing within each
monolayer. Looking once again at Fig.\ \ref{fig1}(c) it is clear that
the monolayer Fermi surface is not capable of supporting the
conventional zero-momentum spin-singlet Cooper pairs: an electron with
crystal momentum $+\bk$ does not have a requisite partner at $-\bk$
with opposite spin that would be needed to form a singlet pair. The 2D
inversion symmetry of the FS,  on the other hand, guaranties that an
equal-spin triplet pair can always be formed from two electrons at
$+\bk$ and $-\bk$, making it a dominant instability of the system.

This is a variant of the argument previously advanced in the context
of genuine altermagnets \cite{Heung2024,Monkman2026} --
but we see that it applies to surface
altermagnets, provided they are composed of weakly coupled
altermagnetic monolayers. Our explicit model calculations support this
intuitive picture and, because it relies only on symmetries and
general principles, we expect the conclusions to be broadly
applicable to the $A$V$_2$$X_2$O family of  materials, regardless of
fine details of chemistry and band structure. In addition, the above insight implies that the triplet
phase should become more stable for weaker interplane coupling -- and
this is indeed confirmed by model calculations \cite{SM}. 

The broader significance of our findings lies in the fact that
spin-triplet superconductors are extremely rare  and the few
known candidates have very low critical temperatures, limiting their
technological usefulness \cite{Kallin2016}. The prospect of Na$_{2-x}$V$_2$Se$_2$O being a
16.3 K triplet superconductor is, therefore, exciting on several
levels. First, chiral $p$-wave is a prototypical topological
superconductor in two dimensions \cite{Read2000}. A thin film of such
a material will feature
a fully gapped bulk with topologically protected edge modes. As
discussed e.g.\ in Ref.\ \cite{Heung2024}, these can be chiral or helical,
depending on the relative helicity of the spin-up and spin-down
condensates. Second, because the two condensates are to leading order
decoupled, the material is capable of carrying spin-polarized
persistent currents, including pure spin supercurrents, which
are of interest for spintronic applications \cite{Monkman2026}.

{\em Conclusions --} Our considerations indicate that the same mechanism
that underlies unconventional spin-triplet superconductivity
theoretically discussed in the context of altermagnetic metals
\cite{Zhu2023,Heung2024,Leraand2025,Monkman2026,Fradkin2026}  is
applicable to the class of metallic antiferromagnets that exhibit
{\em surface} altermagnetism. This has potential applications to 
Na$_{2-x}$V$_2$Se$_2$O which has been reported to exhibit a
magnetic transition near 90 K and superconductivity below 16.3 K \cite{Sun2026SC}. If
this material is proven to be a surface altermagnet with the 
G-type magnetic ordering then our theory unambiguously predicts
equal-spin triplet superconducting order with chiral $p$-wave symmetry.

Very recently, another member of the family 
CsV$_2$Se$_2$O was reported to show a superconducting-like resistivity
downturn below 3 K under compression \cite{Li2026SC}, possibly broadening the
scope of our theory. Clearly, interplay between
superconductivity and magnetism in vanadium
oxychalcogenides is an exciting topic with potential technological
applications and deserves broad attention.

{\em Acknowledgments --} The author is indebted to Alannah Hallas,
Niclas Heinsdorf, Jairo Sinova and Libor \v{S}mejkal  
for stimulating discussions and correspondence. The work
was supported by NSERC and  CIFAR.  The author thanks Aspen Center for
Physics where part of this work was completed.

\bibliography{spin}


\newpage
~
\newpage

\newcommand{\<}{\langle}
\newcommand{\e}{\varepsilon}
\newcommand{\up}{\uparrow}
\newcommand{\down}{\downarrow}
\newcommand{\Up}{\Uparrow}
\newcommand{\Down}{\Downarrow}
\renewcommand{\>}{\rangle}
\renewcommand{\(}{\left(}
\renewcommand{\)}{\right)}
\renewcommand{\[}{\left[}
\renewcommand{\]}{\right]}
\renewcommand{\v}[1]{\boldsymbol{#1}} 
\newcommand{\dslash}{d \hspace{-0.8ex}\rule[1.2ex]{0.8ex}{.1ex}}
\renewcommand{\d}{\partial}
\newcommand{\del}{\nabla}
\renewcommand{\div}{\nabla\cdot}
\newcommand{\curl}{\nabla\times}
\newcommand{\eps}{\epsilon}
\newcommand{\p}{\parallel}
\newcommand{\U}{\mathcal{U}}

\appendix

\setcounter{equation}{0}
\setcounter{figure}{0}
\setcounter{table}{0}
\setcounter{page}{1}
\makeatletter
\renewcommand{\theequation}{S\arabic{equation}}
\renewcommand{\thefigure}{S\arabic{figure}}
\renewcommand{\bibnumfmt}[1]{[S#1]}
\renewcommand{\citenumfont}[1]{S#1}

\onecolumngrid
\begin{center}
  {\bf \large Supplementary Material: Unconventional superconductivity in $A$V$_2$$X_2$O family of surface altermagnets \\ } ~ \\
~ \\
Marcel Franz \\
{\em Department of Physics and Astronomy, and Quantum Matter
  Institute, \\ University of British Columbia, Vancouver, BC, Canada V6T 1Z1}
\end{center}
\vspace{0.5cm}
\twocolumngrid

{\em Mean-field decoupling of the interaction term --}  As an example, we start by
decoupling the second term in the interaction Hamiltonian Eq.\
\eqref{h4} in the spin triplet channel. The relevant terms are
$-V_1(n_{\br\uparrow}  n_{\br'\uparrow} + n_{\br\downarrow}
n_{\br'\downarrow} )$ where $(\br,\br')$ labels a pair of n.n.\ sites
in the same V$_2$O plane. If we define a bond order parameter
\begin{equation}\label{s1}
\Delta^\sigma_{\bdel}=V_1\langle c_{\br\sigma}c_{\br+\bdel\sigma}\rangle
\end{equation}
then we may write the decoupled Hamiltonian as 
\begin{equation}\label{s2}
\cH_I^{\rm MF}=\sum_{\br,\bdel,\sigma}\left[\Delta^\sigma_{\bdel}
  c^\dag_{\br\sigma}c^\dag_{\br+\bdel\sigma}+{\rm h.c.}\right]
+{1\over V_1}\sum_{\br,\bdel,\sigma}|\Delta^\sigma_{\bdel}|^2.
\end{equation}
Here $c^\dag_{\br\sigma}$ denotes a generic creation operator that can 
stand for $a^\dag_{\br\sigma}$ or $b^\dag_{\br\sigma}$ in
different sublattices.

We now specialize to the chiral $p$-wave case by setting
$\Delta^\sigma_{x}=-i\Delta_p/2$ and
$\Delta^\sigma_{y}=\pm\Delta_p/2$. Fourier transforming, we then
obtain 
\begin{equation}
\begin{split}\label{s3}
\cH_I^{\rm MF}=\sum_{\bk,\sigma}\left[\Delta_p(\sin{k_x}\pm i\sin{k_y})
  c^\dag_{\bk\sigma}c^\dag_{-\bk\sigma}+{\rm h.c.}\right]\\
+{2M\over V_1}\sum_{\sigma}|\Delta_p|^2,
\end{split}
\end{equation}
where $M$ denotes the number of lattice sites in each plane. It is easy to
check that $\cH_I^{\rm MF}$ furnishes the in-plane triplet component of the pairing matrix
$\hat\Delta_\bk$ with the $d$-vector given in Table I. 

Combining this with the kinetic term Eq.\ \eqref{h6} one can write
the full BdG Hamiltonian in the form of Eq.\ \eqref{h5}. For
equal-spin triplet order parameter the BdG Hamiltonian is block
diagonal in spin. Each of the $4\times 4$ blocks can be easily
diagonalized and has energy eigenvalues $\pm E_{\bk\alpha}$ given by
Eq.\ \eqref{h6} with $\Delta_\bk=\Delta_p(\sin{k_x}\pm
i\sin{k_y})$.

A completely analogous procedure for the interplane interaction term
$-V_1^z(n_{\br\uparrow}  n_{\br'\downarrow} + n_{\br\uparrow}
n_{\br'\downarrow} )$, where $\br'=\br\pm \hat{z}$, yields the BdG
Hamiltonian with the extended $s$-wave order
$\Delta_\bk=\Delta_s^z\cos{(k_z/2)}$ which furnishes the singlet component of the pairing matrix
$\hat\Delta_\bk$.  The energy eigenvalues are again $\pm E_{\bk\alpha}$ given by
Eq.\ \eqref{h6}. We remark that in this case the spectrum is fully
gapped except for a line node present whenever the Fermi surface
intersects the $k_z=\pi$ plane where $\Delta_\bk$ vanishes.

One can similarly perform the mean-field decoupling in all the other
channels. As an example, we decouple the
on-site interaction represented by the first term in $\cH_I$, that is
$-V_0\sum_\br n_{\br\uparrow}n_{\br\downarrow}$. This can only give a
spin-singlet $s$-wave superconductivity, and the relevant order
parameter is   
\begin{equation}\label{s4}
\Delta_0=V_0\langle c_{\br\uparrow}c_{\br\downarrow}\rangle.
\end{equation}
The interaction term becomes
\begin{equation}\label{s5}
\cH_I^{\rm MF}=\sum_{\br}\left[\Delta_0
  c^\dag_{\br\uparrow}c^\dag_{\br\downarrow}+{\rm h.c.}\right]
+{1\over V_0}\sum_{\br}|\Delta_0|^2,
\end{equation}
or, in the momentum space,

\begin{equation}\label{s6}
\cH_I^{\rm MF}=\sum_{\bk}\left[\Delta_0
  c^\dag_{\bk\uparrow}c^\dag_{-\bk\downarrow}+{\rm h.c.}\right]
+{M\over V_0}|\Delta_0|^2.
\end{equation}

It is easy to check that the above expression corresponds to the $\tau_x$ matrix in the
notation used in the main text. This anticommutes with the first and
the last terms in the kinetic Hamiltonian Eq.\ \eqref{h6}, but commutes
with the middle (magnetic) term. As a result, the spectrum of
excitations of the full BdG Hamiltonian takes a non-BCS  form  $\pm
E_{\bk\alpha}$ with 
\begin{equation}\label{s7}
E_{\bk\alpha}^2=\xi_\bk^2+\eta_\bk^2+g_\bk^2+\Delta_\bk^2+2\alpha\sqrt{\xi_\bk^2(\eta_\bk^2+g_\bk^2)+\eta_\bk^2\Delta_\bk^2}.
\end{equation}
where $\xi_\bk=t_\bk-\mu$ and $\alpha=\pm 1$. This form of the spectrum
is more difficult to analyze in full generality, and it is instructive
to first
consider the case of  $g_\bk=0$, which occurs for $k_z=\pi$. Eq.\ \eqref{s7}
then simplifies to
\begin{equation}\label{s8}
E_{\bk\alpha}=\sqrt{\xi_\bk^2+\Delta_0^2}+\alpha\eta_\bk.
\end{equation}
It is easy to see that the spectrum is gapped on the Fermi surface
only when $|\Delta_0|>|\eta_\bk|$. In realistic materials the SC gap
will be of the order few meV whereas the typical altermagnetic splitting is hundreds of meV,
except near high-symmetry points (along the BZ diagonal in a $d$-wave
altermagnet) where it vanishes. It follows that the spectrum defined by Eq.\
\eqref{s8} is gapless over the majority of the FS where it
intersects with the $k_x=\pi$ plane.

Away from the $k_x=\pi$ plane $g_\bk$ is nonzero  and we find a gapped
spectrum over the rest of the FS. However,
the gap magnitude is suppressed relative to the BCS value
$\Delta_0$. This is illustrated in Fig.\ \ref{s1}(a) where we present the
excitation gap numerically evaluated from  Eq.\ \eqref{s7} along the
intersection of the 3D Fermi surface with several constant-$k_z$
planes. Because of this gap suppression, we expect the on-site
singlet order parameter to be energetically disadvantaged compared
to the fully gapped spectrum and this is indeed borne out in the analysis
of the relevant gap equations below.

{\em Gap equations --} In thermal equilibrium, the system will select
values of SC order parameters that minimize its free energy $F$. The
latter can be computed as $F=-{1\over \beta}\ln{Z}$ where $Z$ is the partition function of the
system. For the triplet phase discussed above we obtain, for each spin
projection,
\begin{equation}\label{s9}
F={N\over 2 V_1}(\Delta_p)^2-{2\over\beta}\sum_{\bk,\alpha}\ln{\left(2\cosh{{1\over 2}\beta E_{\bk\alpha}}\right)},
\end{equation}
and a similar expression for the interplane singlet. Here, $N$ denotes the
number of unit cells in the crystal and the momentum sum extends over
the 3D Brillouin zone. Minimizing this free
energy with respect to the order parameter then yields the general form
of the gap equation listed as Eq.\ \eqref{h9} in the main text. 
\begin{figure}[t]
  \includegraphics[width=8.5cm]{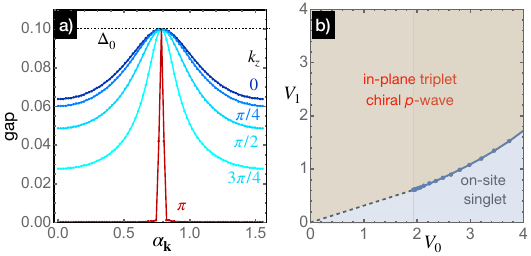}
  \caption{a) Magnitude of the excitation gap in the presence of on-site singlet
    order parameter that follows from the non-BCS spectrum Eq.\
    \eqref{s7}, plotted for several values of $k_z$ as a function of
   the  in-plane angle $\alpha_\bk=\arctan{(k_y/k_x)}$ 
    b) Phase diagram in the $V_0$-$V_1$ plane obtained by
    comparing critical temperatures in the two channels. The dashed
    line is an extrapolation of numerical results. We use $\Delta_0=0.1$, $\eta=0.6$,
    $\mu=-1.14$ and  $g=0.4$.
}
  \label{fig_s1}
\end{figure}

We note that the expression \eqref{s9} for the free energy  is
generally valid whenever the excitation energies of a free-fermion
system come in pairs $\pm E_{\bk\alpha}$.  Therefore, the same approach
can be applied to the case of other order parameters, including the  
on-site singlet discussed above. The specific form
of the gap equation will be more complicated because one needs to take
a derivative of the excitation spectrum given in Eq.\ \eqref{s7}. One
obtains
\begin{equation}\label{s11}
    \Delta_0={V_0\over 4 N}\sum_{\bk,\alpha}{\Delta_0\over
      E_{\bk\alpha}}\left(1+{4\alpha\eta_\bk^2\over E_{\bk+}-E_{\bk-}}
    \right) \tanh{{\beta  E_{\bk\alpha}\over 2}},
\end{equation}
and a similar expression for the in-plane extended
$s$-wave order parameter which  follows by
replacing $\Delta_0\to \Delta_s(\cos{k_x}+\cos{k_y})$  in the
spectrum  Eq.\ \eqref{h9} along with $V_0\to V_1$.
\begin{figure}[t]
  \includegraphics[width=8.5cm]{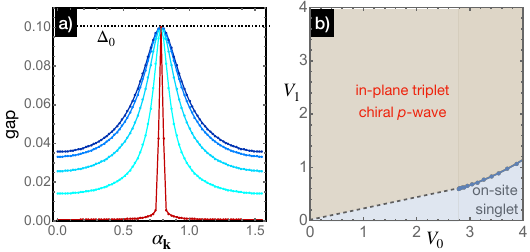}
  \caption{The same as Fig.\ \ref{fig_s1} except for a weaker
    interplane tunneling amplitude $g=0.2$. As compared to Fig.\
    \ref{fig_s1} we observe the on-site gap further 
    reduced away from $\alpha_\bk=\pi/4$, panel (a); as well as
    diminished region occupied by the on-site singlet phase, panel
    (b).   
}
  \label{fig_s2}
\end{figure}

Fig.\ \ref{fig_s1}(b)
shows the phase diagram of the system in the $V_0$-$V_1$ plane
obtained by numerically solving the gap equation \eqref{s11}, its
counterpart for $\Delta_p$ and comparing the relevant critical
temperatures. We observe that once again the phase space is dominated
by the triplet chiral $p$-wave state.

The critical temperatures are most
conveniently calculated by taking the $\Delta\to 0$ limit of the relevant
gap equation. This gives an equation for $T_c$  that only depends on the
underlying normal-metal properties, the symmetry of the order
parameter, and the strength of the attractive 
interaction. As an example, taking the zero-gap limit of Eq.\ \eqref{h10}, one
obtains
\begin{equation}\label{s12}
{1\over {V_1^z}}={1\over 4
              N}\sum_{\bk,\alpha}{\cos^2{(k_z/2)}\over
              \tilde{\epsilon}_{\bk\alpha}}\tanh{{\beta  \tilde{\epsilon}_{\bk\alpha}\over 2}},
\end{equation}
where $\tilde{\epsilon}_{\bk\alpha}=\epsilon_{\bk\alpha}-\mu$ is the
electron band energy referenced to the chemical potential. Eq.\
\eqref{s12} can be solved for $T_c$ by numerical iteration.

{\em Limit of weakly coupled layers --} As mentioned in the main text,
we expect the in-plane spin-triplet pairing to become increasingly more dominant
when the coupling $g$ between the individual V$_2$O layers is reduced. This expectation is
indeed supported by our solution of the relevant gap equations. Fig.\
\ref{fig_s2} shows a clearly reduced phase space occupied by on-site
singlet when the interplane coupling is reduced from $g=0.4$ to $0.2$,
with the triplet taking over a larger portion of the $V_0$-$V_1$ plane.     

\vfill
\eject

\end{document}